\documentclass[twocolumn,showpacs]{revtex4}
\usepackage{graphicx}
\usepackage{subfigure}
\usepackage{color}
\usepackage{amsmath,amssymb}
\begin{document}

\title{Studies of two-solar-mass hybrid stars within the framework of Dyson-Schwinger equations}
\author{Tong Zhao$^{1}$, Shu-Sheng Xu$^{2}$, Yan Yan$^{2}$, Xin-Lian Luo$^{3}$, Xiao-Jun Liu$^{1}$}\email{liuxiaojun@nju.edu.cn}
\author{Hong-Shi Zong$^{2,4,5}$}\email{zonghs@nju.edu.cn}

\address{$^{1}$ Key Laboratory of Modern Acoustics, Department of Physics, Collaborative Innovation Center of Advanced Microstructures, Nanjing University, Nanjing 210093, China}
\address{$^{2}$ Department of Physics, Nanjing University, Nanjing 210093, China}
\address{$^{3}$ Department of Astronomy, Nanjing University, Nanjing 210093, China}
\address{$^{4}$ Joint Center for Particle, Nuclear Physics and Cosmology, Nanjing 210093, China}
\address{$^{5}$ State Key Laboratory of Theoretical Physics, Institute of Theoretical Physics, CAS, Beijing 100190, China}

\begin{abstract}
   In this paper we introduce an equation of state (EOS) of quark matter within the framework of Dyson-Schwinger equations (DSEs) to study the structure of compact stars. The smooth crossover from hadronic matter to quark matter in the hybrid star is studied. We compare different strategies to obtain crossover EOSs and find a new way to construct two-solar-mass hybrid stars with even a relatively soft quark EOS, while earlier works show that the quark EOS should be stiff enough to support a massive hybrid star.

\pacs{12.38.Aw, 12.39.Ba, 14.65.Bt, 97.60.Jd}

\end{abstract}

\maketitle
\section{Introduction}
The equation of state (EOS) of quark matter is crucial for the study of quark stars and compact stars. Once the EOS is obtained, the radius and gravitational mass of a compact star with a given central mass density can be calculated by solving the Tolman-Oppenheimer-Volkoff (TOV) equations. By changing the central density, the mass-radius relationship curve that is used to be compared with the astronomical observations can be drawn. Thus, the astronomical observations can be used to constrain the parameter sets of the EOSs and even rule out some EOSs. However, on the one hand, the perturbative QCD method fails in the low-energy region, and it is very difficult to obtain a reliable EOS from the first principles of QCD. Some approximate methods that incorporate basic features of QCD are adopted to calculate the EOS of quark matter. For example, the Massachusetts Institute of Technology (MIT) bag model and the Nambu-Jona-Lasinio (NJL) model. But all these models have their own weaknesses. The MIT bag model violates chiral symmetry even in the limit of a massless quark. So, there are also some authors developed advanced bag models such as the so called chiral MIT bag model \cite{2}. This model restores the chiral symmetry by introducing the coupling of pion and quark. The NJL model assumes that the interaction between quarks is point-like, so this model is not renormalizable, and it cannot incorporate quark confinement \cite{3}. On the other hand, the observational maximum mass of compact stars is increasing. For example, the $PSR J0348+0432$ was reported to have $2.01\pm0.04$ solar mass in 2013 \cite{4}. Only very stiffen EOS can support such a large maximum mass. Nonetheless, the EOS is often softened when hyperon mixing, deconfined quark matter or kaon condensate is taken into consideration in the core of a compact star. So, F. Ozel even argued that the condensates and unconfined quarks may not exist in the centers of neutron stars \cite{5}.

To solve these problems, people tried to find stiffen EOSs that can support massive quark stars and neutron stars. For instance, in a pure hadronic framework, the introduction of a ``universal 3-body force'' acting on all the baryons or an effective Lagrangian includes quartic terms in the meson fields can make the hyperon‐mixed neutron stars become possible \cite{6, 7}. In MIT bag model, some authors also introduced a density dependent bag parameter to get stiff EOSs \cite{8}. Besides, the smooth crossover from hadronic matter to quark matter is used to construct hybrid stars with high mass recently. Certainly, the order of hadron-quark phase transition at zero temperature is still an open problem and the first-order phase transition (Maxwell or Gibbs construction) is widely adopted even in some recent works \cite{9, 10}, and the exist of the so-called ``mass twins'' in the mass-radius relationship for compact stars is presented as an potential evidence of the first order phase transition \cite{11}. However, we know that the radius of a compact star is very difficult to be determined exactly, so it is not easy to find  ``mass twins'', and there are also some arguments that the ``mass twins'' also exist in the smooth phase transition case \cite{12}. {Besides, lattice QCD (LQCD) studies showed that the transition line for low net baryon is a crossover \cite{13, 131,132,133}, but the results for low temperature and large baryon density are still model dependent. Many people suppose the existence of a critical end point (CEP), but others do not. Actually, on the theoretical side there is still an ambiguity, not only for the location of CEP but also for the existence of CEP~\cite{14}. For example the authors of Ref.~\cite{141} argue that if the vector interaction is strong enough, the transition is a crossover in the whole phase diagram.} Furthermore, treating the point-like hadron as an independent degree of freedom in the Gibbs condition is not fully justified in the transition region because all hadrons are extended objects composed of quarks and gluons. The study on smooth phase transition in hybrid stars is thus also necessary. The picture of a gradual onset of quark degrees of freedom in dense matter associated with the percolation of finite size hadrons has been discussed in some seminal works such as Refs. \cite{15, 16}. Recently, the smooth crossover from hydronic matter to quark matter is used to construct massive hybrid stars \cite{17,18,19}. In these works, based on the percolation picture, the authors utilized different interpolation functions to do some phenomenological studies on the hybrid stars compatible with 2 solar mass. Obviously, the strategies of interpolation are different, but in each paper, the authors just concentrate on a certain interpolation strategy.

The aim of this paper is to find a proper EOS of quark matter to calculate the structure of compact stars and discuss the possibility of two solar mass hybrid star based on the smooth phase transition between the hadronic matter and quark matter. So, we perform a model study on the hybrid star from the point of view of the smooth crossover. For the quark phase, we adopted the EOS with strangeness based on the Dyson-Swinger Equations (DSEs), because the DSEs can simultaneously address both confinement and dynamical chiral symmetry breaking \cite{20, 21}, and it has been applied to hadron physics successfully \cite{22, 23, js, zhao, wx, cui}. We discuss different interpolation strategies to construct hybrid EOSs and compare their influence on the final results, that is to say, the maximum mass of the hybrid stars. We find a new way to construct an EOS that is sufficiently stiff to include stable hybrid stars compatible with two solar mass and our conclusion is an excellent complement to the conclusion in the Ref. \cite{17}. Because it is usually considered that the EOS of quark matter should be stiff enough to support a massive hybrid star compatible with two solar mass before.

This paper is organized as follows. In Sec. II, we introduce the EOS of quark matter. In Sec. III, several ways are tried out to construct hybrid EOS from the point of view of the smooth crossover phase transition. A new way to construct massive hybrid stars even with a soft EOS of quark matter is proposed. Finally, a brief discussion and conclusion is presented in Sec. V.

\section{The EOS of quark matter and the DSEs}
In this section, we briefly introduce our EOS of quark matter based on the DSEs approach. According to Ref. \cite{24}, we start from the zero temperature and zero quark chemical potential version of the quark propagator DSE that reads
\begin{eqnarray}
S(p)^{-1}&=&Z_2\left(ip\!\!\!\slash+Z_mm\right)\nonumber\\
&+&g^2Z_{1F}\int _q\frac{\lambda ^a}{2}\gamma _{\mu }S(q)\Gamma _{\nu }^a(p,q)D_{\mu \nu }(p-q),
\end{eqnarray}

where $S(p)$ is the dressed quark propagator, $Z_2$ is the field-strength renormalization constant, $Z_m$ is the mass renormalization constant where m is the current quark mass, g is the coupling constant, $Z_{1F}$ is the quark-gluon-vertex renormalization constant, $\int _q=\int \frac{d^4q}{(2\pi )^4}$ is a symbol that represents a Poincare invariant regularization of the four-dimensional Euclidean integral, $\lambda ^a$ is the GellMann matrices, $\Gamma _{\nu }^a(p,q)$ is the quark gluon vertex, and $D_{\mu\nu}(p-q)$ is the dressed gluon propagator. The general form of the quark propagator at zero temperature and zero chemical potential reads
\begin{equation}
S(p)^{-1}=ip\!\!\!\slash A(p^2)+B(p^2),
\end{equation}
where $A(p^2)$ and $B(p^2)$ are scalar functions of $p^2$. The renormalization condition is
\begin{equation}
A\left(\zeta ^2\right)=1,
\end{equation}
\begin{equation}
B\left(\zeta ^2\right)=m,
\end{equation}
at sufficiently large space-like $\zeta^2$ \cite{tan,25}. {We choose the renormalization point to be $\zeta =19$ GeV, as used in Refs.~\cite{tan,25}. Another popular choice is $\xi=2~ \rm{GeV}$ in recent study, e.g. Ref.~\cite{sc}. Actually, DSEs is a renormalizable theory, so the choices of $\zeta$ do not affect the physical results.}.  Eq. (1) is the exact result from the first principles of QCD, but we can not solve it directly unless concrete truncations are performed. Rainbow truncation is used in this work, which means a bare vertex is adopted
\begin{equation}
\Gamma _{\nu }^a(p,q)=\frac{\lambda ^a}{2}\gamma _{\upsilon },
\end{equation}
and the Qin-Chang gluon propagator model \cite{26} is specified by a choice for the effective interaction in Landau gauge. Finally, one can obtain the following equations for the two dressing functions $A(p^2)$ and $B(p^2)$,
\begin{equation}
\begin{aligned}
A\left(p^2\right)=Z_2+\frac{4}{3p^2}\int _q\frac{\mathcal{G}\left(k^2\right)}{k^2}\frac{A\left(q^2\right)}{q^2A^2\left(q^2\right)+B^2\left(q^2\right)}\\ \times \left(p\cdot q+2\frac{(k\cdot p)(k\cdot q)}{k^2}\right),
\end{aligned}
\end{equation}
\begin{equation}
B\left(p^2\right)=Z_4m+4\int _q\frac{\mathcal{G}\left(k^2\right)}{k^2}\frac{B\left(q^2\right)}{q^2A^2\left(q^2\right)+B^2\left(q^2\right)}.
\end{equation}
The coupled Eqs. (6) and (7) can then be solved by direct iteration. The DSE solution for the dressed quark propagator can be well fitted with 3 pairs of complex conjugate poles with the representation \cite{27}
\begin{equation}
S(p)=\sum _{n=1}^3 \left(\frac{z_n}{\text{ip\!\!\!\slash}+m_n}+\frac{z_n^*}{\text{ip\!\!\!\slash}+m_n^*}\right).
\end{equation}
We fit u, d and s quarks respectively, where the following requirement, that the quark propagator in the region of ultraviolet should tend to the free quark propagator, is employed
\begin{equation}
\sum _{k=1}^3 \left(z_k+z_k^*\right)=1.
\end{equation}
In this paper, we choose a new parameter set that can not only fit the propagator well but also lead to a stiff EOS. The corresponding parameters for u, d quarks are
\begin{equation}
\begin{array}{ll}
m_1=285+121i\; {~\rm MeV},&z_1=0.308+0.618i,\\
m_2=-1120+59i\; {~\rm MeV},&z_2=0.111-0.193i,\\
m_3=1214+429i\; {~\rm MeV},&z_3=0.081,
\end{array}
\end{equation}
and for s quark,
\begin{equation}
\begin{array}{ll}
m_1=509+236i\; {~\rm MeV},&z_1=0.327+0.449i,\\
m_2=-1166+616i\; {~\rm MeV},&z_2=0.101+0.00178i,\\
m_3=1572+660i\; {~\rm MeV},&z_3=0.072+0.017i,
\end{array}
\end{equation}
If we compare these parameters with that in the Ref. \cite{24}, we will find that the smaller real parts of $m_1$ and $z_1$ make the EOS stiffer. Then, the propagator at zero temperature and zero chemical potential case can be generalized to nonzero temperature and chemical potential case by the following replacement \cite{28}
\begin{equation}
p^4\rightarrow \tilde{\omega }_n=(2n+1)\text{$\pi $T}+\text{i$\mu $},
\end{equation}
which is widely used within rainbow truncation of DSEs in thermal and dense QCD, and $\tilde{\omega }_n$ is the Matsubara frequencies. Hence, the quark propagator at nonzero T and $\mu$ is
\begin{equation}
S\left(\overset{\rightharpoonup }{p},\tilde{\omega }_n\right)=\sum _{k=1}^3 \left(\frac{z_k}{\text{ip\!\!\!\slash}+\text{i$\gamma $}_4\tilde{\omega }_n+m_k}+\frac{z_k^*}{\text{ip\!\!\!\slash}+\text{i$\gamma $}_4\tilde{\omega }_n+m_k^*}\right).
\end{equation}
Then, the well-known formula for the quark number density is \cite{28, 29}
\begin{equation}
\rho (\mu ,T)=-N_cN_fT\sum _{n=-\infty }^{+\infty } \int \frac{d^3\overset{\rightharpoonup }{p}}{(2\pi )^3}\text{tr}\left[S\left(\overset{\rightharpoonup }{p},\tilde{\omega }_n\right)\gamma _4\right].
\end{equation}
After a series calculations and Taking the limit $T\rightarrow 0$, the final result of the quark number density in zero temperature can be obtained:
\begin{equation}
\begin{aligned}
\rho (\mu ,T=0)\\=\frac{N_cN_f}{3\pi ^2}\sum _{k=1}^3 \left(z_k+z_k^*\right)\theta \left(\mu -\mu _k^0\right)\left(\mu ^2-\frac{d_k^2}{4\mu ^2}-c_k\right){}^{\frac{3}{2}}
\end{aligned}
\end{equation}
where $\mu _k^0=\left|\text{Re}\left(m_k\right)\right|$ and $c_k, d_k$ are defined by $m_k^2=c_k+d_ki$. The quark number density dependence on $\mu$ at $T=0$ for s quark is displayed in Fig. 1, and from this figure we can see that the quark number density of s quark turns to nonzero at $\mu _c=520$ MeV, which is smaller than the Ref. \cite{24} because of our new parameter set. So, this EOS is more appropriate for the quark matter with strangeness.
 \begin{figure}
    \begin{center}
      \includegraphics[width=0.5\textwidth]{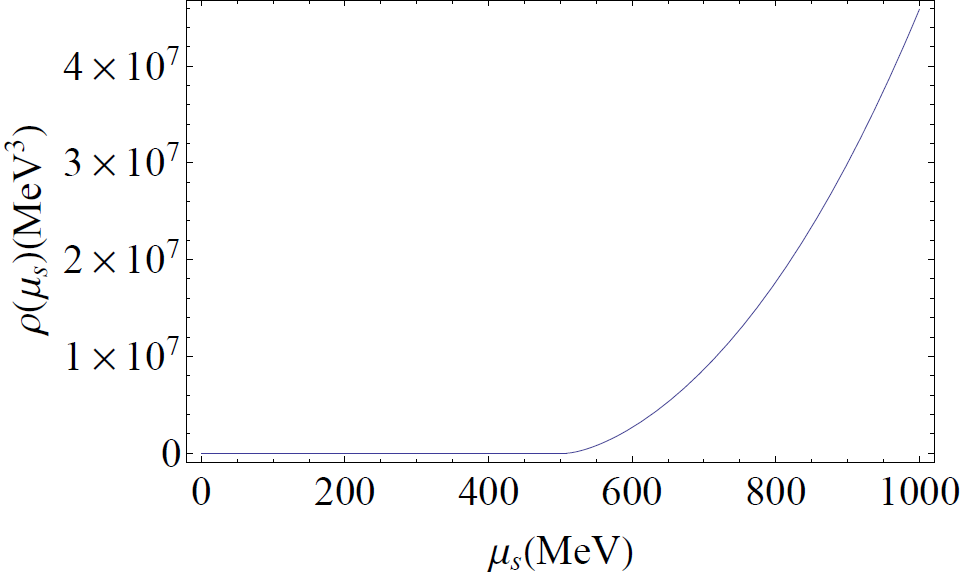} \caption{The quark number density of s quark}
   \end{center}
\end{figure}
Finally, we take the chemical equilibrium and electric charge neutrality conditions into consideration to constrain different quark chemical potentials. The conditions read
\begin{equation}
\mu _d =\mu _u +\mu _e,
\end{equation}
\begin{equation}
\mu _s =\mu _u +\mu _e,
\end{equation}
\begin{equation}
\frac{2}{3}\rho _u-\frac{1}{3}\rho _d-\frac{1}{3}\rho _s-\rho _e=0.
\end{equation}
Then there is only one independent chemical potential due to these constraint. We choose $\mu _u$ in this work. For definite chemical potential of quark, the EOS of QCD at $T=0$ reads \cite{29, 30}
\begin{equation}
P(\mu )=P(\mu =0)+\int _0^{\mu }\rho (\mu ')d\mu ',
\end{equation}
and the relation between the energy density and the pressure of the corresponding system is \cite{31, 32}
\begin{equation}
\epsilon =-P+\sum _i \mu _i\rho _i,
\end{equation}
where $\mu _i$ and $\rho _i$ represents the chemical potential and the particle number density for each component in the system.

\section{The smooth crossover and the structure of hybrid stars}
As usually, one can obtain the structure of a bare quark star by directly integrating the Tolman-Oppenheimer-Volkoff (TOV) equations with the equation of state:
\begin{equation}
\begin{aligned}
\frac{dP(r)}{dr}=-\frac{G(\varepsilon +P)\left(M+4\pi r^3P\right)}{r(r-2GM)},\\
\frac{dM(r)}{dr}=4\pi r^2\varepsilon .
\end{aligned}
\end{equation}

However, in the Eq. (19), we can see that the $P(\mu =0)$ is only a $\mu$-independent constant. It is the pressure of the vacuum. Although a method to calculate the $P(\mu =0)$ self-consistently is given in the Ref. \cite{28}, there is actually no model independent way to calculate the reliably $P(\mu =0)$ from the first principles of the QCD. In this paper, analogous to the MIT bag model, we reconsider the effect of this term and think that there exists negative pressure at zero chemical potential in the vacuum which manifests the confinement of QCD. {Namely, we identify $P(\mu =0)$ with -B, where B is the vacuum bag constant}. In this paper we take it as a phenomenological parameter only. Our EOS for nuclear matter is the APR EOS with $A18+\delta v+UIX^*$ interaction in Ref. \cite{33}. This EOS is based on Argonne v18 two-body potential and the Urbana IX three-body interaction, includes charge neutrality and Beta-equilibrium. The $\delta v$ indicates the inclusion of relativistic corrections. For simplicity, the APR model includes only nucleonic degrees of freedom, and does not take into account hyperons, whose interactions with nucleons and among themselves are not well determined. Then, we choose $B=(110{~\rm MeV})^4$ to ensure that the energy density of the quark matter should always be higher than the hadronic matter in the low density region. But with this value of B, the maximum mass of a pure quark star is just about 1.65 solar mass. If we choose a smaller B, the EOS will become stiffer and the maximum mass will become larger, but such a small B is unreasonable. So, we just try to find a way to construct massive hybrid stars in the point of view of smooth crossover phase transition with such a soft EOS for quark matter.
In order to facilitate analysis and compare with the first order phase transition, the pressure as a function of the baryon chemical potential of both the quark matter and the hadronic matter is showed in the Fig. 2.
 \begin{figure}
    \begin{center}
      \includegraphics[width=0.5\textwidth]{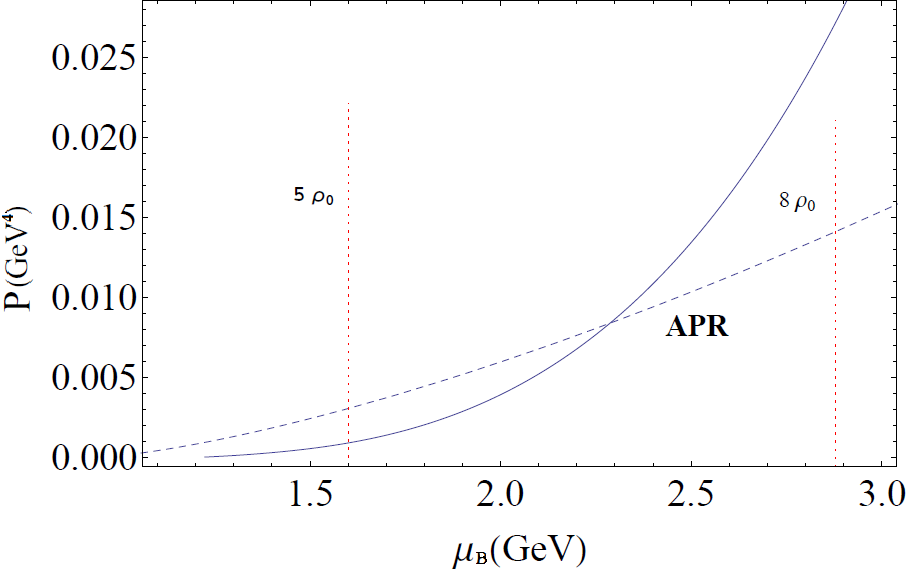} \caption{The pressure as a function of the baryon chemical potential. The dashed line is the EOS of hadronic matter, the solid line represents the EOS of quark matter. The vertical dashed lines show the corresponding baryon number density of the hadronic matter. The baryon number density is scaled by the baryon number density of normal nuclear matter $\rho _0 =0.17fm^{-3}$}
   \end{center}
\end{figure}
From Fig. 2 we can see that the intersection in the $P-\mu _B$ plane is just the first order transition point. However the corresponding baryon number density at this point is over 5 times of the normal nuclear matter($\rho _0 =0.17fm^{-3}$). The hadronic EOS is usually not reliable at such a high density. Besides, the system must be strongly interacting in the transition region, so that it can be described neither by an extrapolation of the hadronic EOS from the low-density side nor by an extrapolation of the quark EOS from the high-density side \cite{34}. And we should point that it is not a special result in our hadronic and quark EOS. Because the LQCD shows that that even in relatively large region of chemical potential, QGP is still strongly interacting, So the pressure of the quark EOS tends slowly to the free quark gas in the $P-\mu _B$ plane. That means the pressure increasing slowly in the $P-\mu _B$ plane, and the intersection between quark matter and hadronic matter usually occurs at high chemical potential and high number density if we choose a modern model to calculate the EOS for quark matter rather than a MIT bag or a perturbative model.
Now, we start to construct EOS for hybrid stars in the point of view of the smooth crossover. From the Refs. \cite{17,18,19} we can see that the strategy is not unique, but it should meet the requirements of physics. Refer to the Ref. \cite{17}, the so called ``3-window-model'', we construct the hybrid EOS from the $P-\rho _B$ plane. The pressure as a function of the baryon number density is showed in the Fig. 3. The shadow region is the possible crossover region that is between the $2\rho _0$ and $4\rho _0$.
 \begin{figure}
    \begin{center}
      \includegraphics[width=0.5\textwidth]{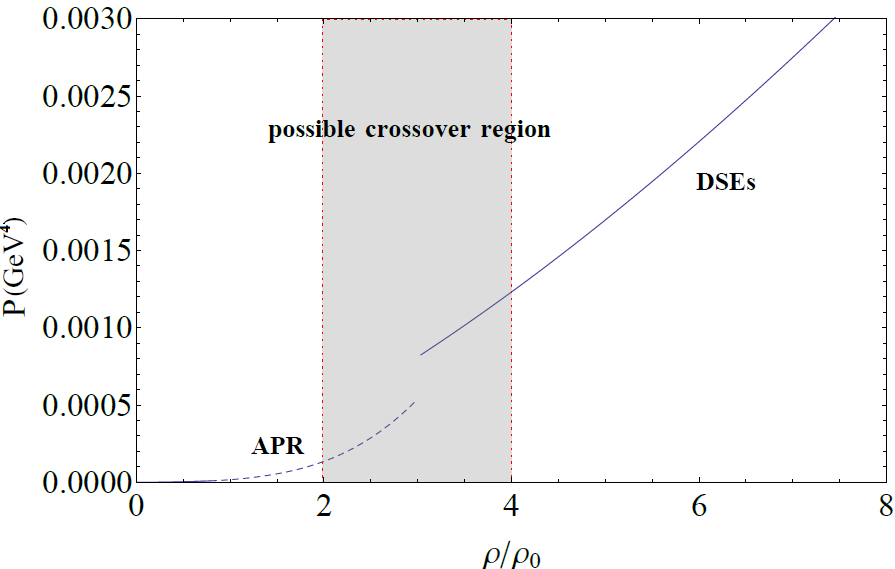} \caption{The pressure as a function of the baryon number density. The shadow region is possible crossover region, the solid line is the EOS of quark matter and the dashed line is the APR EOS. The x-axis is the baryon number density scaled by nuclear number density and the y-axis is the pressure}
   \end{center}
\end{figure}
We adopt the interpolation function in the Ref. \cite{17} to make a smooth connection between the two EOS in the $P-\rho _B$ plane:
\begin{equation}
P=P_H\times f_-+P_Q\times f_+,
\end{equation}
\begin{equation}
f_{\pm }=\frac{1}{2}\left(1\pm \tanh \left(\frac{\rho -\bar{\rho }}{\Gamma }\right)\right),
\end{equation}
where $P_H$ and $P_Q$ are the pressure in the hadronic matter and that in the quark matter,The window $\bar{\rho }-\Gamma \lesssim \rho \lesssim \bar{\rho }+\Gamma$ characterizes the crossover region in which both hadrons and quarks are strongly interacting, so that neither pure hadronic EOS nor pure quark EOS are reliable. From the thermodynamical relation, $P=\rho ^2\partial (\epsilon /\rho )/\partial \rho$, we obtain
\begin{equation}
\epsilon (\rho )=\epsilon _H(\rho )f_-(\rho )+\epsilon _Q(\rho )f_+(\rho )+\Delta \epsilon
\end{equation}
\begin{equation}
\Delta \epsilon =\rho \int _{\bar{\rho }}^{\rho }\left(\epsilon _H(\rho ')-\epsilon _Q(\rho ')\right)\frac{g(\rho ')}{\rho '}d\rho '
\end{equation}
with $g(\rho )=\frac{2}{\Gamma }\left(e^X+e^{-X}\right)^{-2}$ and $X=\frac{\rho -\bar{\rho }}{\Gamma }$. Here $\epsilon _H (\epsilon _Q)$ is the energy density obtained from Hadronic EOS and quark EOS. $\Delta \epsilon$ is an extra term which guarantees the thermodynamic consistency.  Correspondingly, if we start from the energy density and deduce the pressure(This is called $\epsilon -interpolation$ in the Ref. \cite{17}), there will be a $\Delta P$ term in the expression of the pressure. But we find the $\Delta P$ term often leads to a fluctuation in the hybrid EOS and makes the sound velocity ($v_s=\sqrt{\frac{\text{dP}}{\text{d$\epsilon $}}}$) become larger than light. This is obviously unreasonable, so we don't try this.
The final result of this hybrid EOS is showed in the Fig. 4, compared with the quark EOS and hadronic EOS.
 \begin{figure}
    \begin{center}
      \includegraphics[width=0.5\textwidth]{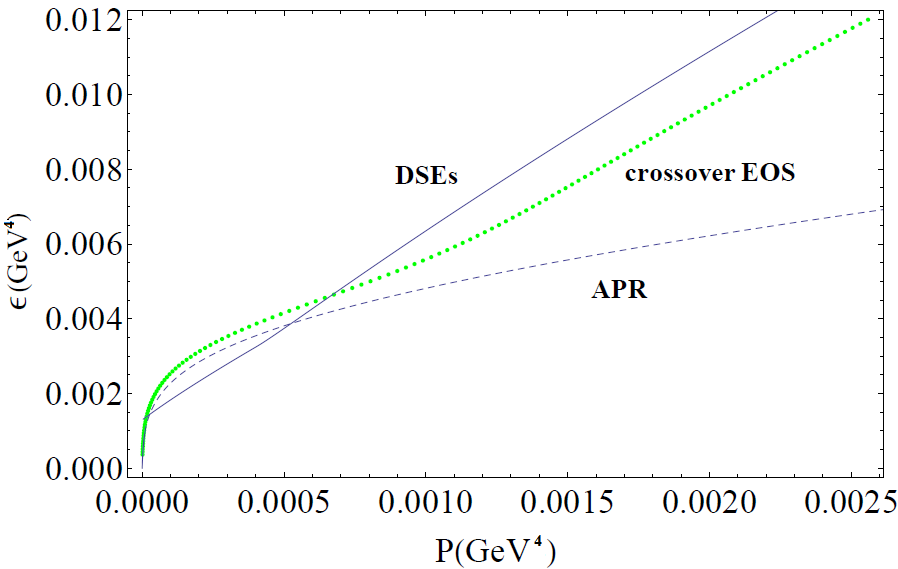} \caption{The interpolation EOS(green dot line) compared with the quark EOS(solid line) and hadronic EOS(dashed line).}
   \end{center}
\end{figure}
From the Fig. 4, the relation between energy density and pressure, we can evaluate the stiffness of the EOS easily. The APR EOS has a well-known stiff tail in the large density region. So, in the  Fig. 4 we can see that when the pressure and energy density are large, the slope of the APR EOS becomes much lower than that of the DSEs EOS. While, in the low density region, the quark EOS is often stiffer than the hadronic EOS. The hybrid EOS based on first order phase transition equals to the hadronic EOS in low density region, and in the large density region it equals to the quark EOS. So, the hybrid EOS based on first order phase transition is often softer than both the hadronic EOS and the quark EOS. However, if we construct the hybrid EOS from smooth crossover, the EOS will be influenced by the quark EOS in the low density region, and be influenced by the hadronic EOS in the large density region. So, the result will be different.  Unfortunately, our hybrid EOS is still very soft if we utilize this interpolation strategy, because our quark EOS is soft. This is in accordance with the conclusion in the Ref. \cite{17} that the maximum mass of hybrid stars can exceed two solar mass only if the quark matter has a stiff equation of state and the crossover takes place at around three times the normal nuclear matter density.
However, we find another way to construct two solar mass hybrid stars with even a soft quark EOS. We apply the same interpolation function to the $P-\mu _B$ plane and then calculate energy density from thermodynamical relations. The pressure, energy density and the EOS are showed in the Fig. 5, Fig. 6 and Fig. 7. {The chemical potential in the crossover region is between 1.4 GeV to 3.4 GeV because the first order phase transition point is at around 2.4 GeV. In order to quantify the stiffness of EOSs, the sound velocities are showed in the Fig. 8 and Fig. 9. A comparison can be found in Ref. [21]. The EOS obtained by the interpolation from the $P-\rho _B$ plane is soft while the EOS obtained by the interpolation from the $P-\mu _B$ is much stiffer. From Fig. 8 we can see that the sound velocity is always smaller than 0.4 times of the velocity of light while the sound velocity in Fig. 9 is much larger.}
 \begin{figure}
    \begin{center}
      \includegraphics[width=0.5\textwidth]{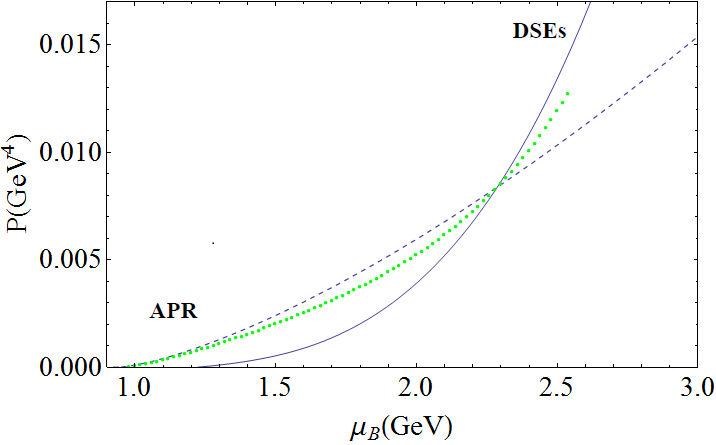} \caption{The pressure as a function of the baryon chemical potential. The green dot line shows the pressure of interpolation EOS, the solid line is the pressure of the quark matter and the pressure of the hadronic matter is the dashed line.}
   \end{center}
\end{figure}
 \begin{figure}
    \begin{center}
      \includegraphics[width=0.5\textwidth]{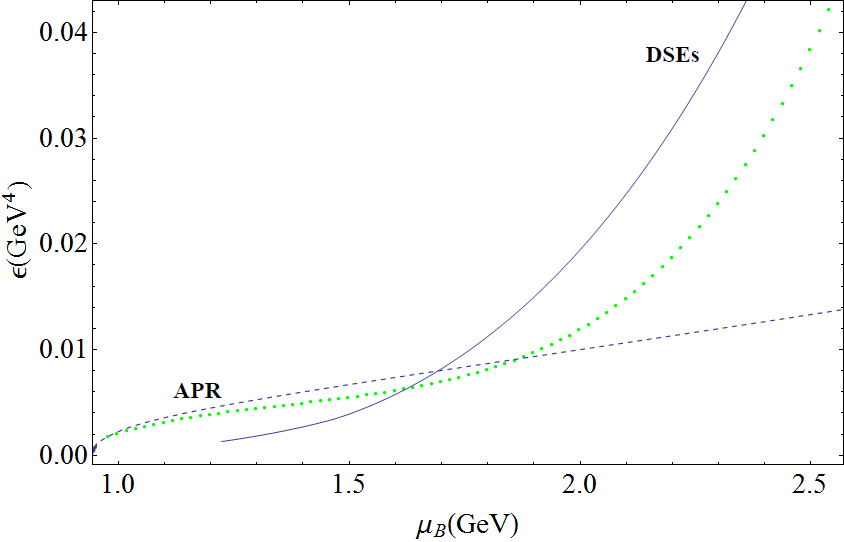} \caption{The energy density as a function of the baryon chemical potential. The green dot line shows the energy density of the interpolation EOS, the solid line is the energy density of the quark matter and the energy density of the hadronic matter is the dashed line.}
   \end{center}
\end{figure}
 \begin{figure}
    \begin{center}
      \includegraphics[width=0.5\textwidth]{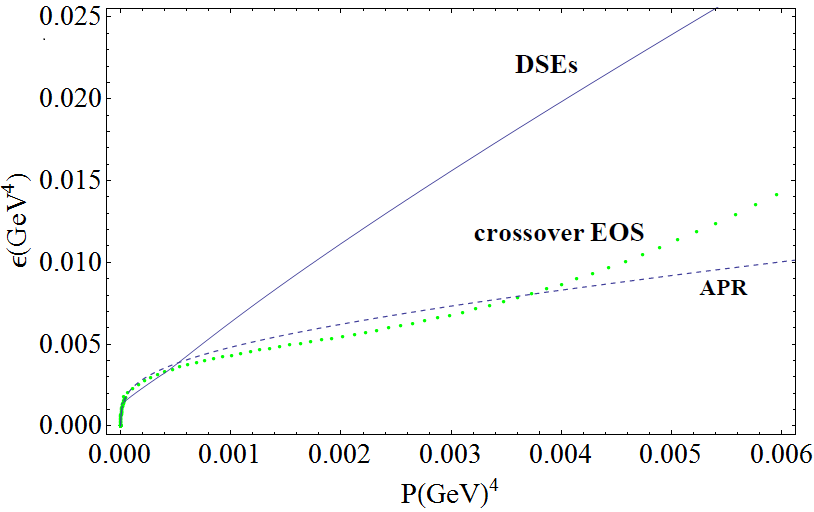} \caption{The interpolation EOS(green dot line) compared with the quark EOS(solid line) and hadronic EOS(dashed line).}
   \end{center}
\end{figure}

\begin{figure}
    \begin{center}
      \includegraphics[width=0.5\textwidth]{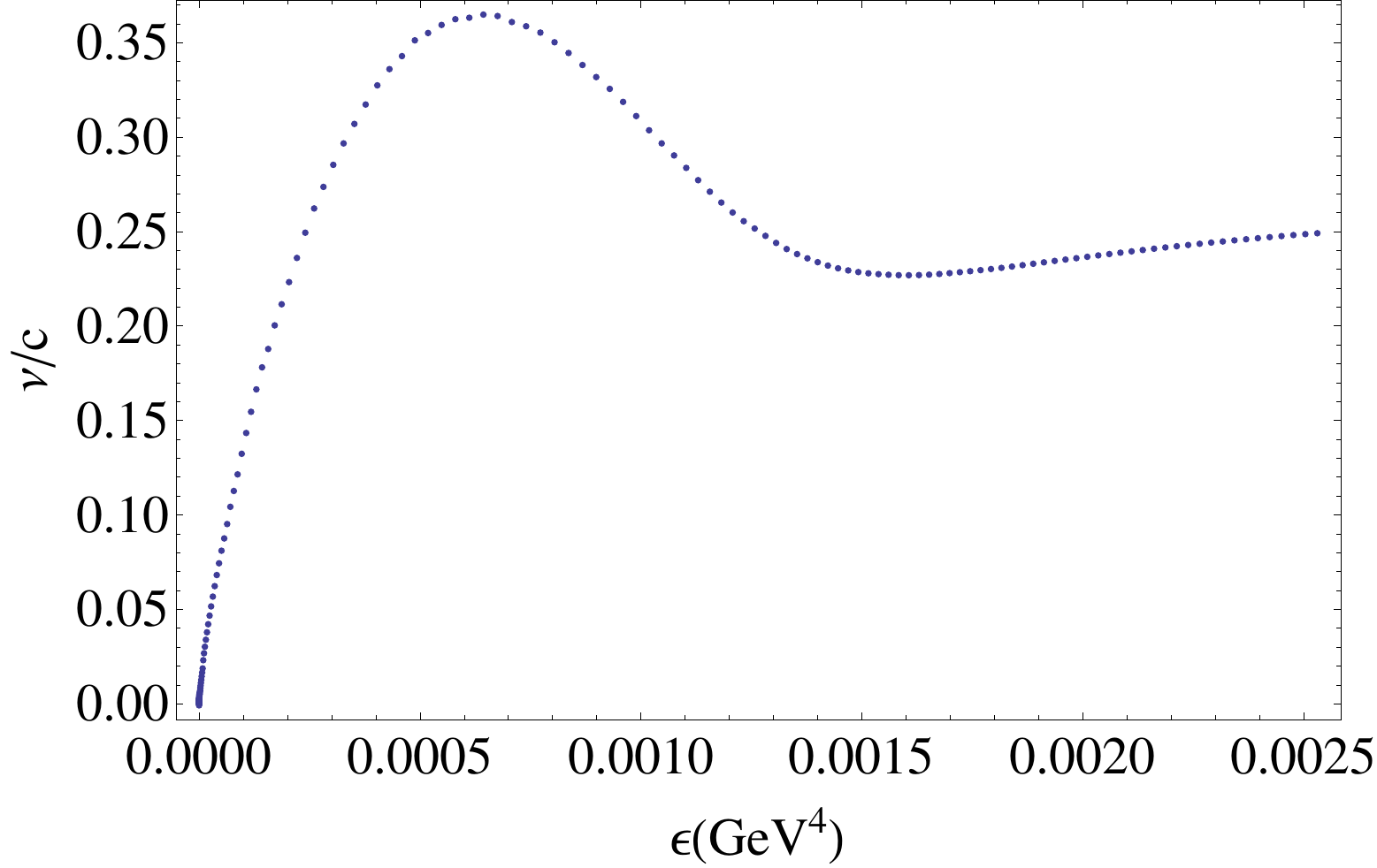} \caption{The sound velocity of the hybrid EOS obtained by the interpolation from the $P-\rho _B$ plane}
   \end{center}
\end{figure}
\begin{figure}
    \begin{center}
      \includegraphics[width=0.5\textwidth]{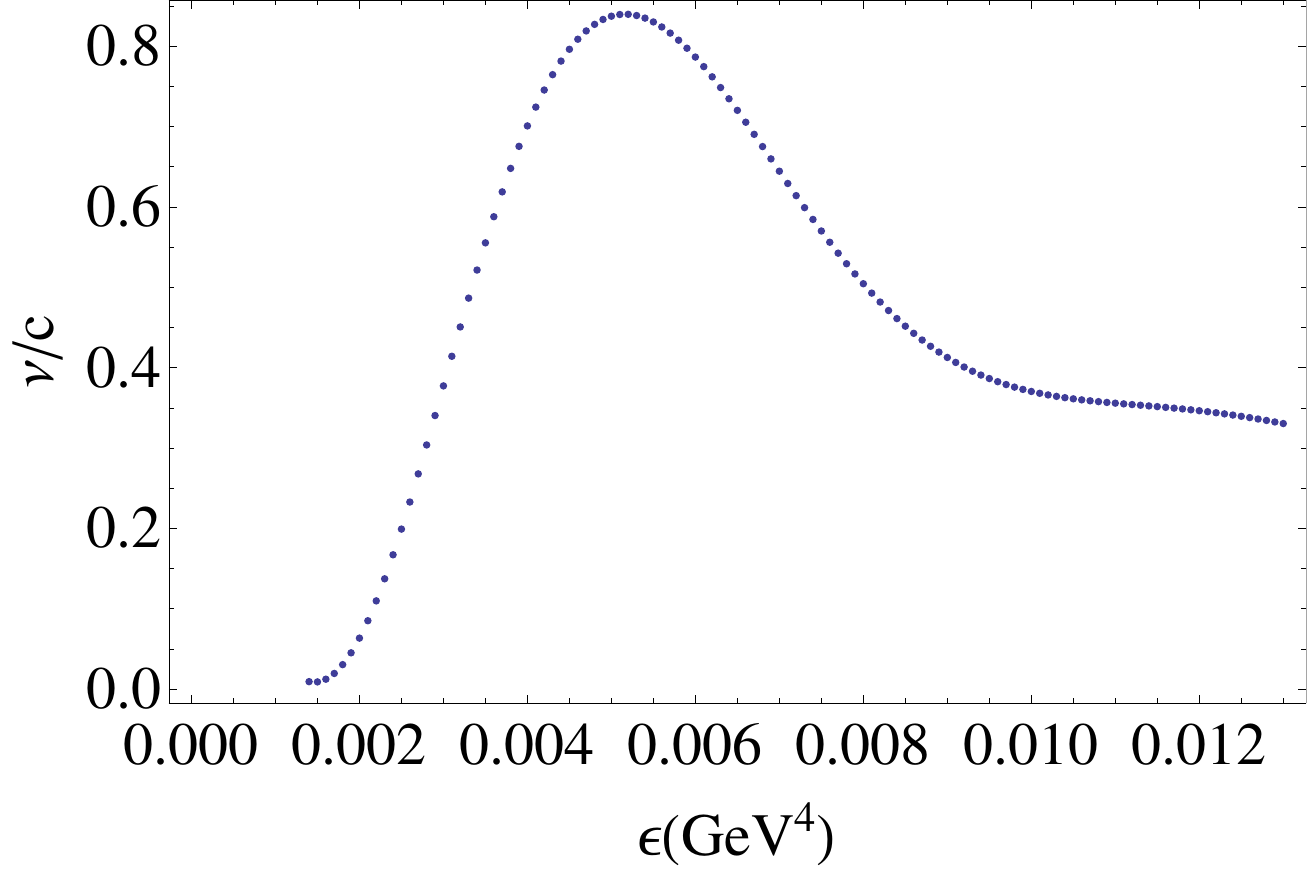} \caption{The sound velocity of the hybrid EOS obtained by the interpolation from the $P-\mu _B$ plane}
   \end{center}
\end{figure}

In the {$P-\mu_B$} plane, we can clearly find the difference between the phase transition of smooth crossover and Maxwell construction. Although the pressure tends to quark matter in the high chemical potential region and tends to the hadronic matter in the low chemical potential region, it is different from both the quark matter and the hadronic matter in the crossover region. Finally, by integrating the TOV equations with the EOS, we get the mass-radius relation of the hybrid stars. It is showed in the Fig. 10. The maximum mass is {2.35} times the solar mass. There are also other interpolation functions to make the interpolation EOS. For example, a polynomial function is used in the Ref. \cite{19}. But we find that different interpolation functions cannot make appreciable difference. If we adopt the polynomial function to construct the EOS, the difference of the maximum mass is less than 5 percent. This result is apprehensible, because the interpolation functions should be smooth at the boundaries of the interpolating interval. Thus, the values of different interpolation functions in the crossover region do not differ widely.
 \begin{figure}
    \begin{center}
      \includegraphics[width=0.5\textwidth]{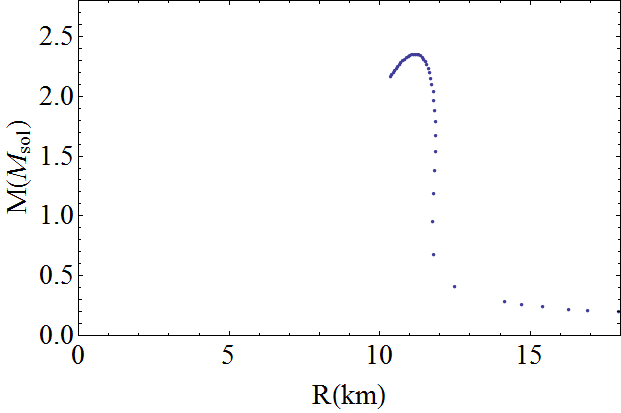} \caption{The m-r relation of the hybrid stars based on the smooth crossover EOS. The maximum mass is over two solar mass}
   \end{center}
\end{figure}

\section{Discussion}
In this paper we introduce our quark EOS with 3-flavor quark based on the framework of DSEs to calculate the structure of hybrid star. For the hadronic phase we adopted the APR EOS and we calculate the mass-radius relationship of the hybrid stars in the point of view of the smooth crossover phase transition from hadronic matter to quark matter. The common belief is that the EOS for quark phase should be very stiff to construct a two solar mass hybrid star. For example, in the Ref. \cite{17}, the authors performed the interpolation in the $P-\rho$ plane and claimed that no matter what kind of hadronic EOS is adopted, the maximum mass of neutron stars can exceed 2 solar mass on condition that the crossover takes place at around three times the normal nuclear matter density and the quark matter is strongly interacting in the crossover region and has stiff EOS. Nonetheless, we find that if we start from the $P-\mu$ plane, the interpolation function can generate a stiff EOS and finally construct a massive hybrid star compatible with two solar mass, even though our quark EOS is relatively soft.

{For simplicity, the hyperons are not included because the interaction of hyperons with nucleons and among themselves are not well determined. The consideration of hyperon will soften the EOS for hadronic matter,  but in our result, the maximum mass of the hybrid stars is larger than the maximum mass of both the pure quark stars and the pure hadronic stars. So, to what extent the influence  of hyperons will be is model dependent and need further study.}

\acknowledgments
This work is supported in part by the National Natural Science Foundation of China (under Grants No. 11275097, No. 11475085, and No. 11535005), and the National Basic Research Program of China (under Grant 2012CB921504).

\end{document}